\documentstyle[12pt,aaspp4]{article}

\begin{document}

\title{Density Variations over Subparsec Scales in Diffuse Molecular Gas}

\author{K. Pan\altaffilmark{1,2}, 
S. R. Federman\altaffilmark{1,2}, and
D. E. Welty\altaffilmark{3,4}}

\altaffiltext{1}{Department of Physics and Astronomy, University of Toledo,
    Toledo, OH 43606.}
\altaffiltext{2}{Guest Observer, McDonald Observatory, University of Texas 
at Austin.}
\altaffiltext{3}{Astronomy and Astrophysics Center, University of Chicago, 
5640 South Ellis Ave., Chicago, IL 60637.}
\altaffiltext{4}{Visiting Observer, Kitt Peak National Observatory, National
 Optical Astronomy Observatories, which is operated by AURA, Inc., under 
cooperative agreement with the National Science Foundation.}

\begin{abstract}
We present high-resolution observations of interstellar CN, CH, CH$^{+}$, 
\ion{Ca}{1}, and \ion{Ca}{2} absorption lines toward the multiple star systems
HD206267 and HD217035. 
Substantial variations in CN absorption are observed among three sight lines of 
HD206267, which are separated by distances of order 10,000 AU; 
smaller differences are seen for CH, CH$^+$, and \ion{Ca}{1}. 
Gas densities for individual velocity components are inferred from a 
chemical model, independent of assumptions about cloud shape.  
While the component densities can differ by factors of 5.0 between adjacent 
sightlines,  the densities are always less than 5000 cm$^{-3}$. 
Calculations show that the derived density contrasts are not sensitive to the
temperature or reaction rates used in the chemical model.  
A large difference in the CH$^{+}$ profiles (a factor of 2 in column density)
is seen in the lower density gas toward HD217035. 
\end{abstract}


\section{Introduction}

Evidence for pervasive subparsec-scale structure in the diffuse 
interstellar medium (ISM) has been accumulating over the past three decades 
through radio and optical/ultraviolet 
observations. Observations of \ion{H}{1} 21 cm 
absorption toward high-velocity pulsars (Frail et al. 1994) and toward 
extended extragalactic radio sources 
(Faison \& Goss 2001) reveal variations 
in optical depth on scales of 10 to 100 AU. Radio observations of absorption 
from H$_{2}$CO and OH toward extragalactic sources (Moore \& Marscher 1995) 
also show changes in optical depth over time.  
Multi-epoch optical measurements provide another probe of 
these small-scale structures (e.g., Lauroesch, Meyer, \& Blades 2000; Price, 
Crawford, \& Barlow 2000; Crawford et al. 2000; Welty \& Fitzpatrick 
2001).  On intermediate scales of 10$^{2}$ to 10$^{5}$ AU, 
optical/ultraviolet observations of interstellar absorption lines 
(mainly from the trace species \ion{Na}{1} and \ion{K}{1}) toward individual 
members of binary or multiple star systems (e.g., Meyer \& Blades 1996; 
Watson \& Meyer 1996; Lauroesch et al. 1998; Lauroesch \& Meyer 1999--LM99) 
suggest ubiquitous structure.  Structure on these scales 
is also implied by variations of \ion{Na}{1} and \ion{K}{1} absorption 
profiles toward members of clusters (e.g., Langer, Prosser, \& Sneden 1990; 
Bates et al. 1995) and toward bright extended cores of globular clusters 
(Meyer \& Lauroesch 1999; Andrews, Meyer, \& Lauroesch 2001).  In this 
$Letter$, we present optical spectra at a resolution of $\sim$ 1.7
km s$^{-1}$ which reveal differences in molecular profiles across two 
multiple star systems.

While it is well established that the column densities of some atomic 
(\ion{H}{1}, \ion{Na}{1}, \ion{K}{1}) and  molecular (H$_{2}$CO, 
OH) species vary on scales of 10 to 10$^{5}$ AU in diffuse sight lines,
detailed information on the physical conditions causing the 
variations is generally not available. A key question posed by the 
observations is whether the subparsec structure is caused by density 
variations (e.g., Frail et al. 1994; LM99; Crawford 
et al. 2000), by fluctuations in ionization equilibrium 
(LM99; Welty \& Fitzpatrick 2001), 
by the geometric structure of clouds (Heiles 1997), or by something else. 
A crucial point is to determine accurate physical 
conditions for the gas showing  variations in column density. 
As a step in this direction, 
we present high resolution absorption spectra of 
CN, CH, CH$^{+}$, \ion{Ca}{1} and \ion{Ca}{2}, 
discuss the observed velocity component structure, and derive gas 
densities of components based on a simplified chemical model for CH, 
C$_{2}$, and CN (Federman et al. 1994; Knauth et al. 2001). We also 
investigate the sensitivity of the derived densities on cloud temperature and 
chemical reaction rates.  

\section{Observations and Data Reduction}

High resolution spectra were obtained for two multiple star systems 
in the active star-forming regions Cep OB2 and OB3 (Table 1). 
Most of the observations were acquired in 2000 
September with the 2d-coud\'{e} spectrograph on the 2.7 m 
telescope at McDonald Observatory (Tull et al. 1995). A 
cross-disperser and the availability of a thin Textronix 
2048$\times$2048 CCD made it possible to get CN, CH, CH$^{+}$, 
\ion{Ca}{1}, and \ion{Ca}{2} features in a single exposure.
In order to achieve a final 
signal-to-noise (S/N) ratio greater than 50, five 30-min exposures were 
obtained for each target. Comparison spectra from a Th-Ar hollow cathode lamp
were taken every two hours, while biases and flat fields were acquired 
each night. Dark frames were also obtained as a check on the amount of 
thermal noise from the cooled CCD. The setup yielded a spectral 
resolution of $\sim$ 1.7 km s$^{-1}$ which was determined from the 
widths of the Th-Ar lines. The star HD206267A was observed
in 1995 with the 0.9 m Coud\'{e} Feed Telescope at 
Kitt Peak National Observatory. Spectra for \ion{K}{1}, 
\ion{Ca}{2}, CN, and CH were obtained using camera 6, 
the echelle grating, and the F3KB CCD, at resolutions 
of about 1.3 km s$^{-1}$. Comparison of \ion{K}{1} spectra of HD206267A 
obtained in 1995, 1998, and 2001 reveal no obvious temporal variation.

The raw images were reduced in the usual manner with the NOAO IRAF 
echelle data reduction package. The individual frames were 
first bias-subtracted. Cosmic-ray hits and scattered light were removed 
from each stellar image and flat field. The flat fields were divided into the 
stellar images to account for differences in pixel-to-pixel sensitivity. 
Since each echelle order was a few pixels wide, the orders were summed 
across this width. Then all extracted one-dimensional spectra were 
calibrated in wavelength using the Th-Ar comparison spectra. After 
correcting for Doppler motion, individual spectra for each sight line 
were combined. Finally, the combined spectra were normalized to unity using
low-order polynomials. Final spectra appear in Figures 1 and 
2; their S/N ratios range from 50 to 150. 

Component structure is clearly seen in the spectra. The equivalent 
width, $W_\lambda$, for each velocity component was determined by Gaussian
fit. We attempted to use the same velocity structure to fit all the absorption 
features in all lines of sight in each stellar system.  While very good fits 
for HD217035A/B and HD206267A/C were obtained with essentially the same 
component structures,
a slightly different component 
structure was required to fit the spectra of HD206267D. For each 
sightline, however, the same velocity components are used for all species.
Toward HD206267C, for example, CN, CH, CH$^+$, \ion{Ca}{1}, and \ion{Ca}{2} all
have components at about $-$3.1 km s$^{-1}$ and 0.9 km s$^{-1}$. The adopted
velocity components for each species are indicated in Fig. 1 and 2.

Each value of $W_\lambda$ for the CH and CN components was converted into a 
column density using curves of growth.  The Doppler parameter ($b$-value) 
was set at 1.0 km s$^{-1}$ (Federman et al. 1994), a value consistent with 
the measured line widths. The inferred optical depths at
line center were found to be modest ($\leq$1). Therefore, our column densities
are not susceptible to unresolved structure. Moreover, even for a $b$-value 
of 0.5 km s$^{-1}$, the column density ratio between sight lines for a given 
velocity component --- from which the density contrasts are 
inferred --- changes by only 15\%. 
When only the R(0) line of CN at 3874.61 \AA\ was detected for a component, 
the column density $N$(CN) for the component was set equal
to 1.5 times the value from R(0), as would be expected for an excitation 
temperature of about 2.7 K. If a CH component was detected but with no 
corresponding CN absorption, a 3-$\sigma$ detection limit is given. 
Table 2 lists the column densities for the HD206267 
system, where we find significant variations in these quantities; 
a complete listing of results will appear elsewhere. 

\section{Analysis and Results}

\subsection{Spectral Appearance}

Two members of multiple star system HD206267 (C/D)
were observed with the same instrumental setup. 
As shown in Figure 1, there are striking differences in the CN, CH, 
CH$^{+}$, and \ion{Ca}{1} absorption profiles between the two sight lines. 
In addition to differences in component strengths, the \ion{Ca}{1} profiles 
appear slightly ``shifted'' with respect to one 
another. Because CH$^{+}$ and \ion{Ca}{1} absorption lines occur 
in the same echelle order, because the wavelength solutions for the two stars 
are the same, and because the CH$^{+}$ profiles 
toward the two stars match so well in velocity space, we conclude that the 
``shift'' cannot be attributed to wavelength calibration. Since a 
slightly different component structure was required
to get good fits for the CN and CH$^{+}$ profiles for the two sight lines, 
the ``shift'' is more likely the result of slightly different component 
structure.

For comparison, the slightly higher resolution KPNO spectra of CN, CH, and 
\ion{Ca}{2} toward HD206267A
were smoothed to the effective resolution of the spectra for
HD206267C/D. While Gaussian fits show that the interstellar absorption
toward A and C exhibits the same component structure,
large variations in line strength are found 
for both CN and CH.  
Significant variations are also seen between the spectra toward C and D for all 
observed species 
except \ion{Ca}{2}.  It is noteworthy that CN, which 
is a good tracer of dense gas (e.g., Federman et al. 1994), 
shows the largest variations.

Figure 2 shows significant differences in the CH$^{+}$ absorption
profiles for the two members of the HD217035 system, but the CH, 
\ion{Ca}{1}, and \ion{Ca}{2} profiles are very similar. Absorption from
CN was not detected in either sight line. The CH$^{+}$ 
column densities (based on a curve of growth with $b$ of 2.5 km s$^{-1}$)
toward B are greater than the 
corresponding ones toward A by factors of 2.5, 1.5, 1.3, and 1.1 for 
the components at $-$14.3, 5.5, $-$10.5, and 
$-$6.7 km s$^{-1}$, respectively. In contrast, the 
CH and \ion{Ca}{1} column densities for all 
components show no variations between the two sight lines at the 10\% level. 

\subsection{Chemical Modeling}

The significance of the line profile variations in Figures 1 and 2 
and the column density variations seen in Table 2 is revealed through a 
simplified chemical model (Federman et al. 1994; Knauth et al. 2001),  
which provides estimates for the gas density in each component. 
The chemistry of the simple molecular species, CH, C$_{2}$, and CN, is 
relatively well understood (van Dishoeck \& Black 1986;
Federman et al. 1994); we use the formalism of  
Federman et al. (1994) in our analysis.

The fractional abundance of C$^+$ in diffuse gas,  
$x$(C$^+$) $\sim$ 1.4 $\times$ 10$^{-4}$ (Meyer et al. 1997), is smaller by
a factor of 1.8
than the value adopted by us in earlier studies (Federman \& Huntress 1989; 
Federman et al. 1994).  In order to preserve the rates needed to reproduce 
observations of C$_2$ and CN, the rate coefficients,
$k_1$ (for C$^{+}$~+ CH~$\rightarrow$~C$^{+}_{2}$~+~H) and $k_5$ 
(for C$^{+}$~+~NH~$\rightarrow$~all products), were correspondingly increased 
to 5.4$\times10^{-10}$ and 5.0$\times10^{-10}$ cm$^{3}$ s$^{-1}$, respectively. 
In addition, the grain optical depth 
corresponding to the transition from C$^+$ to CO was 
increased from 2.0 to 2.3; otherwise, gas densities somewhat larger than those
typically found in diffuse molecular gas are obtained.  
All other input parameters remain unchanged. 
The derived gas densities for all 
components are listed in Table 2.  
Uncertainties are based on the $\pm$1--$\sigma$ ranges in observed
CH and CN columns. For the components at $-3.1$,  $-0.7$ and 0.9 km s$^{-1}$, 
the chemical analysis provides evidence for factors of 1.5 to more
than 5 
changes in density as the cause of the CN profile variations seen in 
Fig. 1, but in no case do we infer densities as large as 5000 cm$^{-3}$.
While the CN excitation temperature can be used as an independent 
measure of the (electron) density (Black \& van Dishoeck 1991), our data 
are not of sufficient precision for such an analysis.

\section{Discussion}

Pervasive subparsec structure in the diffuse ISM, revealed by variations 
in atomic and molecular column densities, is beyond doubt.  
While other studies attempted to extract information on 
changes in physical condition related to these 
column density variations, the analyses were often based on 
the {\it ad hoc} assumption of a spherical cloud to 
determine the gas density structure. 
Moreover, the estimated densities are very sensitive to the 
temperature, which is generally not well known (Heiles 1997; Crawford et al. 
2000).  For instance, Crawford et al. (2000) used the column density ratio,
$N$(\ion{K}{1})/$N$(\ion{Na}{1}), to estimate the physical conditions in 
relatively dense clouds. Toward HD81188, an increase in that ratio by a 
factor of 1.9, on a scale of $\sim$ 15 AU, could be explained either by 
increasing gas density by a factor of 5 or by decreasing temperature  
by a factor of 10.  More likely, it may be due to some combination 
of density increase and temperature decrease.  In many optical studies, an 
important unknown factor is the fractional ionization,  which is needed to 
estimate $N$(H) from trace neutral species such as \ion{Na}{1} 
(Lauroesch et al. 1998; Welty \& Fitzpatrick 2001).

In our chemical analysis, the density is 
determined essentially by a column density ratio, $N$(CN)/$N$(CH), 
which reduces influences of cloud shape on density structure.  
For example, the $-$3.1 km s$^{-1}$ component 
toward HD206267C and the $-$0.7 km s$^{-1}$ one toward HD206267D have similar
$N$(CN)/$N$(CH) --- and thus comparable 
gas densities --- even though their CN and CH column densities differ by a 
factor of about 3. Our derived gas density also is not sensitive to 
temperature. For instance, calculated densities for velocity 
components toward A and C increase 
by only 3 to 8\% when the temperature of 40 K is halved.  The change in gas 
density $ratios$ for a given velocity 
component in the two sight lines is even smaller, 
1--2\%, when the temperature is halved. 

Calculations also show that the gas density ratio for a given velocity
component is not overly sensitive to the reaction rates in the chemical model. 
The rate coefficients $k_{1}$ and $k_{5}$ produce the 
largest effects on the gas density; decreasing either 
increases the calculated density. 
When both rate coefficients are decreased by 80\%, 
the derived gas densities 
toward HD206267A and HD206267C increase by 50--80\%, but the gas 
density ratios for individual velocity components in the two sightlines
change by only 4--8\%. The greatest change 
in this density ratio (from 1.58 to 1.72) 
is found for the $-0.7$ km s$^{-1}$ component.
We conclude that this chemical analysis is a good 
approach for deriving physical properties of small scale structure in 
diffuse gas clouds with detectable amounts of CN. 

The HD206267 multiple star system is located in a region of recent, 
active star formation. LM99 studied \ion{K}{1} 
variations toward members of this system, and estimated typical 
gas density ratios of 1.1--2.7 for velocity components toward 
A and C. [Here we adopt $N$(\ion{K}{1}) $\propto$
$n^{2.7}$ instead of $n^{1.7}$ to take into account the additional factor of 
density in $N$(\ion{K}{2}).] The density 
ratios derived from CH and CN range from 1.5 to greater than 5.0.  The lower 
density ratios found by LM99 presumably can be attributed to \ion{K}{1} 
absorption probing a more extended volume of the cloud than CN absorption.  
The larger volume likely smooths out the large density
variations in CN-rich gas.  This picture is consistent with our \ion{Ca}{2} 
observations, which show no significant profile variations, probably because 
\ion{Ca}{2} absorption occurs over an even more extended region 
(Crinklaw, Federman, \& Joseph 1994; Welty, Morton, \& Hobbs 1996).  
The present investigation also supports the conjecture of LM99 
that much of the variation toward this system may 
be in dense gas. Table 
2 shows that the gas density variations among sight lines of the system are 
detected only in components with $v_{LSR}$ from $-$4.0 to 1.0 
km s$^{-1}$, with no obvious large density variation for the 4.3 km s$^{-1}$ 
component. The CO emission-line maps of Patel et al. (1998) 
indicate that CO-rich material in 
these lines of sight only appears within a velocity range 
of $v_{LSR}$ $\sim$ $-$5.0 km s$^{-1}$ to $\sim$ 2.0 km s$^{-1}$ --- the 
range in which we see significant variations. Furthermore, using the \ion{K}{1}
results of LM99 with the trends determined by Welty and Hobbs (2001) for
H$_{tot}$, we find H$_{tot}$ of $\approx$1$\times$$10^{21}$ cm$^{-2}$ for these
components and only 
about 4$\times$$10^{20}$ cm$^{-2}$ for the $+4.3$ km s$^{-1}$ one.
Our gas densities then suggest cloud thicknesses of about 1 pc along the
lines of sight, similar to the sizes of the CO cloudlets seen in emission 
(Patel et al. 1998).

Variations in CN and CH column densities were not detected 
toward the HD217035 system.  Since only upper limits for CN are available for 
these sight lines, the density must be relatively low.  The variations in the 
CH$^+$ profiles toward both systems may help to elucidate the uncertain 
chemistry for this molecule. 

In summary, our observations of CN and CH absorption show that 
the diffuse molecular clouds toward the multiple star 
system HD206267 exhibit significant density structure. The gas density 
varies by factors greater than 5.0 over a scale of $\sim$ 10,000 AU. 
The present approach to studying small-scale structure in diffuse clouds 
provides reliable density contrasts, but it is 
limited to relatively dense molecule-rich clouds. 

\acknowledgments

It is our pleasure to thank the support staff of McDonald Observatory 
and KPNO, especially David Doss at McDonald Observatory.  This work 
was supported by NASA grants NAG5-4957 and NAG5-8961 
to the University of Toledo.

\clearpage

\begin{deluxetable}{lcccrrc} 
\tablecolumns{7}
\tablenum{1}
\tablecaption{Stellar Data}
\tablewidth{0pt}

\tablehead{
System & $V$ & $E$($B-V$) & $D$ &
\multicolumn{2}{c}{Separation} & Reference \\ \cline{5-6}
 & (mag) & (mag) & (pc) & (arcsec) & (AU) & }
\startdata
HD206267A/C/D & 5.7/8.0/7.9 & 0.52/0.57/0.42 & $615\pm$15& 
AC: 11.8 & 7,250 & 1,2 \\
               &             &                &           & CD: 31.6 & 
               19,410 & \\
               &             &                &           & AD: 19.9 & 
               12,220 & \\
HD217035A/B   & 8.6/8.6     & 0.76/0.76      & 725       & 1.9       & 
1,380  & 3,4 \\ 
\enddata     
\tablerefs{(1) ESA 1997; (2) de Zeeuw et al. 1999; 
(3) Crawford \& Barnes 1970; (4) Dommanget \& Nys 1994.}
\end{deluxetable}

\begin{deluxetable}{lcccccc} 
\tablecolumns{7}
\tablenum{2}
\tablecaption{Column Densities and Densities for 
the HD206267 System}
\tablewidth{0pt}

\tablehead{
Sight Line & Quantity & \multicolumn{5}{c}{ $v_{LSR}$ (km s$^{-1}$) } \\
\cline{3-7}
 & & -4.1 & -3.1 & -0.7 & 0.9 & 4.3 }
\startdata
HD206267A& $N$(CN) (10$^{12}$ cm$^{-2}$)& $\ldots$ & 7.0$\pm$0.2 & 
0.9$\pm$0.2 & $<$ 0.2 & $<$ 0.2 \\
         & $N$(CH) (10$^{12}$ cm$^{-2}$)& $\ldots$ & 14.8$\pm$0.5& 
         3.6$\pm$0.3 & 10.8$\pm$0.5 & 3.5$\pm$0.3 \\
         & $n$ (cm$^{-3}$)              & $\ldots$ & 1500$\pm$110 & 
         700$\pm$120 & $<$ 60          & $<$ 170 \\
HD206267C& $N$(CN) (10$^{12}$ cm$^{-2}$)& $\ldots$ & 4.9$\pm$0.3 & 
0.9$\pm$0.2 &  1.8$\pm$0.3 & $<$ 0.3 \\
         & $N$(CH) (10$^{12}$ cm$^{-2}$)& $\ldots$ & 12.5$\pm$0.8& 
         2.4$\pm$0.5 & 10.7$\pm$0.8 & 3.8$\pm$0.6 \\
         & $n$ (cm$^{-3}$)              & $\ldots$ & 1000$\pm$200& 
         1100$\pm$500 & 400$\pm$90   &  $<$ 180 \\
HD206267D& $N$(CN) (10$^{12}$ cm$^{-2}$)& 1.3$\pm$0.2 & $\ldots$ &  
1.4$\pm$0.2 & $<$ 0.3 & $<$ 0.3 \\
         & $N$(CH) (10$^{12}$ cm$^{-2}$)& 6.7$\pm$0.4 & $\ldots$ & 
         4.5$\pm$0.4 & 8.0$\pm$0.4 &7.1$\pm$0.5 \\
         & $n$ (cm$^{-3}$)              & 600$\pm$130   & $\ldots$ & 
         950$\pm$220 &  $<$ 130 &  $<$ 150 \\ \hline
\enddata
\end{deluxetable}

\clearpage

\setcounter{figure}{0}
\begin{figure}[p]
\begin{center}\plotfiddle{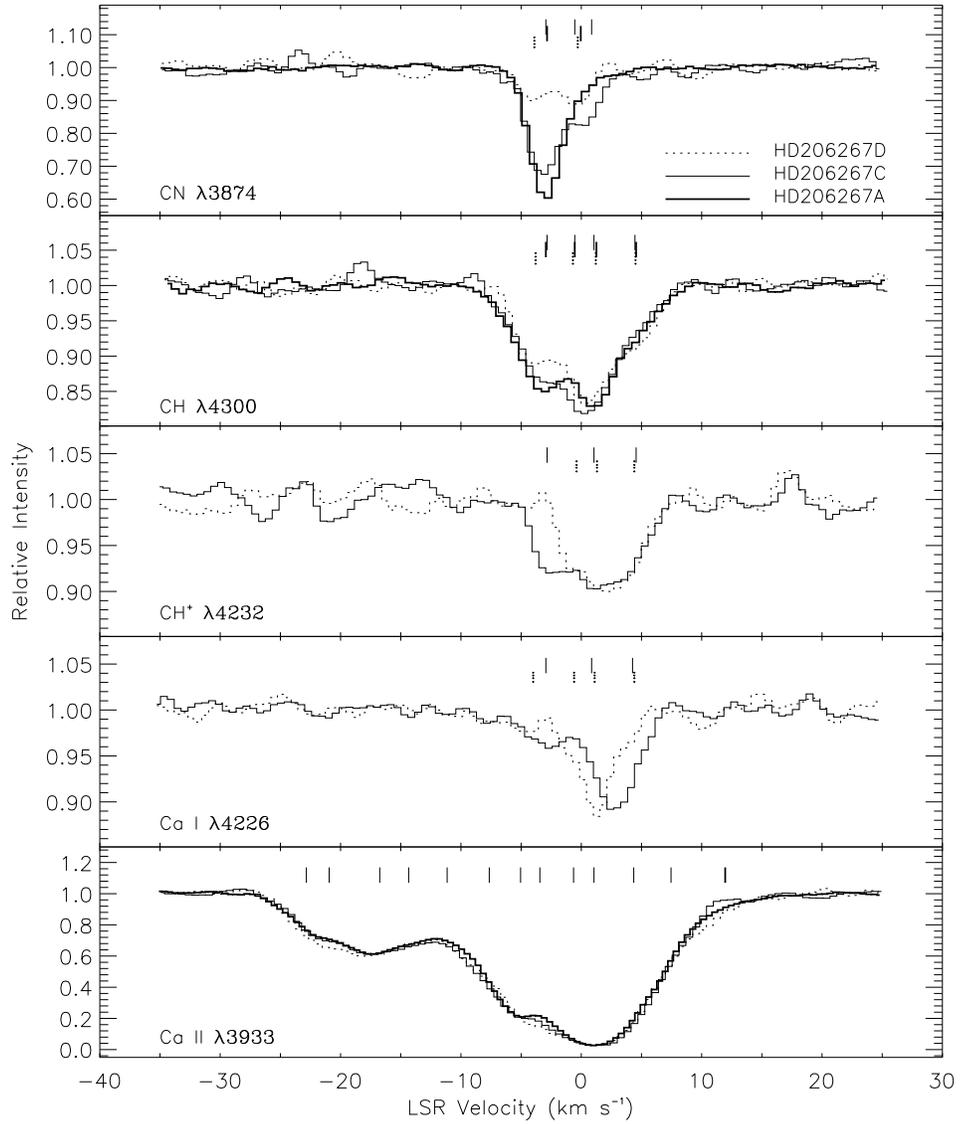}{5.0in}{0}{70}{70}{-190}{-60}
\vspace{0.1in}
\caption{Absorption from interstellar CN, CH, CH$^+$, 
\ion{Ca}{1}, and \ion{Ca}{2} in the spectra of HD206267A (thick solid lines), 
HD206267C (solid lines) and 
HD206267D (dotted lines).  The observed line is indicated.  Note the relative 
intensity scales for the panels differ. Hash marks above each spectrum show
the component structure derived from Gaussian fits.}
\end{center}
\end{figure}

\clearpage

\setcounter{figure}{1}
\begin{figure}[p]
\begin{center}\plotfiddle{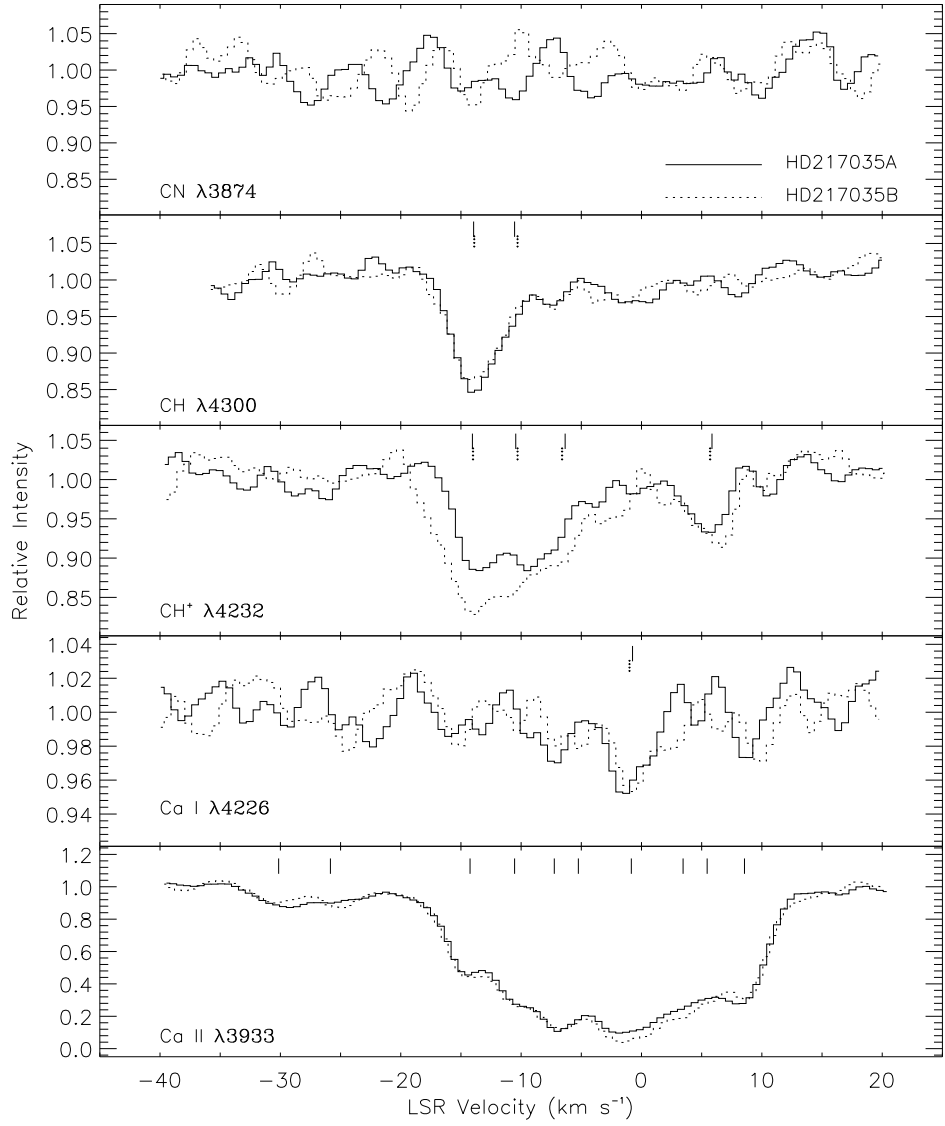}{5.0in}{0}{70}{70}{-190}{-60}
\vspace{0.1in}
\caption{Same as Fig. 1 for HD217035A (solid lines) and 
HD217035B (dotted lines).}
\end{center}
\end{figure}

\end{document}